# Gate-controlled skyrmion and domain wall chirality

Charles-Elie Fillion[1], Johanna Fischer[1], Raj Kumar[1], Aymen Fassatoui[2], Stefania Pizzini[2], Laurent Ranno[2], Djoudi Ourdani[3], Mohamed Belmeguenai[3], Yves Roussigné[3], Salim-Mourad Chérif[3], Stéphane Auffret[1], Isabelle Joumard[1], Olivier Boulle[1], Gilles Gaudin[1], Liliana Buda-Prejbeanu[1], Claire Baraduc[1] & Hélène Béa[1,4] ✉

Magnetic skyrmions are localized chiral spin textures, which offer great promise to store and process information at the nanoscale. In the presence of asymmetric exchange interactions, their chirality, which governs their dynamics, is generally considered as an intrinsic parameter set during the sample deposition. In this work, we experimentally demonstrate that a gate voltage can control this key parameter. We probe the chirality of skyrmions and chiral domain walls by observing the direction of their current-induced motion and show that a gate voltage can reverse it. This local and dynamical reversal of the chirality is due to a sign inversion of the interfacial Dzyaloshinskii-Moriya interaction that we attribute to ionic migration of oxygen under gate voltage. Micromagnetic simulations show that the chirality reversal is a continuous transformation, in which the skyrmion is conserved. This control of chirality with 2–3 V gate voltage can be used for skyrmion-based logic devices, yielding new functionalities.

Magnetic skyrmions are spin-swirling, topologically nontrivial spin textures that hold promise for next-generation spintronic devices[1–5]. Their nanometric size and efficient manipulation by electric current[6] would enable high storage density and fast computational operations. In thin multilayered ferromagnetic films, skyrmions are characterized by circular, homochiral Néel domain walls (DWs), which are stabilized by interfacial Dzyaloshinskii–Moriya interaction (iDMI)[7,8]. The sign of the iDMI constant $D$ sets the preferred chirality of the Néel DW[9]. With our conventions, Néel DW adopts a right-handed or clockwise (CW) chirality for $D < 0$ and a left-handed or counterclockwise (CCW) chirality for $D > 0$.

Besides, chirality plays a key role in the DW dynamics driven by spin–orbit torques[9–12]. In heavy-metal/ferromagnet/metal-oxide (HM/FM/MO$_x$) trilayers, a charge current flowing in the HM layer generates a transverse spin current due to the spin Hall effect whose angular momentum is tranferred to the FM magnetization[13]. The resulting spin–orbit torque moves DWs and skyrmions in a direction that depends on their chirality and on the sign of the spin Hall angle (SHA). It promotes the spin–orbit torque driven motion as an efficient tool to locally probe the chirality of domain walls, and thus the iDMI sign. For instance, a HM underlayer with negative SHA, such as Ta[14,15], induces a motion of CW DWs along the current density whereas CCW DWs move along the electron flow[15].

It is generally considered that iDMI is an intrinsic parameter set during the sample deposition. The effective iDMI in HM/FM/MO$_x$ trilayers is the sum of the contributions originating from the two FM interfaces and may be adjusted by varying the FM thickness[16,17], removing the metal-oxide[18], changing the type of HM[19,20] or the oxidation state at the FM/MO$_x$ interface[21,22]. It has even been shown that tuning the oxidation state of the FM/MO$_x$ interface can invert the iDMI sign[23]. These techniques for controlling iDMI and thus DW chirality were limited to materials engineering until the very recent experimental demonstration of a dynamical and reversible control of chirality by chemisorption[24]. However, this method is not local and necessitated a complex experimental setup. Controlling iDMI using a local and application-compatible external excitation on full solid-state devices would thus open a novel degree of freedom to efficiently manipulate chiral spin textures such as magnetic skyrmions[21].

[1]Université Grenoble Alpes, CEA, CNRS, Spintec, 38000 Grenoble, France. [2]Université Grenoble Alpes, CNRS, Néel Institute, Grenoble, France. [3]Laboratoire des Sciences des Procédés et des Matériaux (LSPM), Villetaneuse, France. [4]Institut Universitaire de France (IUF), Paris, France. ✉e-mail: helene.bea@cea.fr

In particular, gate voltage control of interfacial magnetic properties[25–27] offers a promising, low power and versatile technique to achieve both a local and dynamical control of iDMI. It is well-established that a gate voltage can modify the charge distribution and tune the oxidation state at the $FM/MO_x$ interface, both mechanisms leading to changes in interfacial magnetic anisotropy[28]. The strongest effect associated with non-volatility has been explained as driven by $O^{2-}$ ionic migration towards the interface or away from it, depending on the voltage polarity[29–31]. Such ionic migration is already exploited as a mechanism for resistive switching in anionic metal-oxide memristor devices[32]. This tuning of interfacial magnetic anisotropy has allowed controlling with a gate the creation and annihilation of skyrmions[33–36]. Furthermore, it was demonstrated that the iDMI amplitude is reversibly tunable with a gate voltage[27,37,38] due to its interfacial nature. The possibility to electrically reverse the sign of the iDMI would provide a versatile and reversible control of skyrmion chirality, which could considerably improve their all-electrical, low-power manipulation.

In this article, we demonstrate experimentally that a gate voltage induces a local and dynamical reversal of skyrmion chirality in Ta/FeCoB/$TaO_x$ trilayer. Moreover, we show a similar effect on chiral DWs in a labyrinthine magnetic state, confirming our findings on magnetic skyrmions. We show that regardless of the initial DW chirality, which is controlled by the initial oxidation level at the FeCoB/$TaO_x$ interface, a gate voltage with appropriate polarity is able to switch chirality back and forth in a persistent way. This reversal is attributed to ionic migration, and thus oxidation or reduction of the FeCoB/$TaO_x$ interface, by the gate voltage, which results in the inversion of iDMI sign. Finally, using micromagnetic simulations we show that an adiabatic chirality reversal of a nanometer size skyrmion is possible in Co-based sample. The internal structure of the DW evolves continuously from one chirality to the other without skyrmion annihilation when iDMI is vanishing.

## Results
### Skyrmion chirality reversal with gate voltage
A schematic representation of the Ta(3)/FeCoB(1.2)/$TaO_x$(0.85–1) trilayer (nominal thicknesses in nm), with 20 nm $ZrO_2$ oxide and transparent Indium Tin Oxide (ITO) electrode (see *Methods*), is shown in Fig. 1a. The oxidation step after the top-Ta wedge deposition induces an oxidation gradient at the top interface (see Fig. 1b). This gradient induces a sign crossover of iDMI, as directly measured by Brillouin Light Scattering (BLS, see Fig. 1c and *Methods*). Under zero applied magnetic field, demagnetization of the sample occurs and labyrinthine domains are formed. As observed by polar-Magneto-Optical Kerr-Effect (p-MOKE) microscope, the current-induced motion (CIM) of these DWs is inverted at the $D = 0$ position interpolated from BLS measurements (see dashed line in Fig. 1b, c and Supplementary section I). We will thus further use CIM as a tool to probe iDMI sign where usual state-of-the art quantitative techniques, such as BLS, cannot resolve such small iDMI values (typically around ± 10 μJ/$m^2$ in the regions of interest marked by the triangle and star, see Fig. 1b, c and Supplementary section VI).

In the area close to the iDMI sign crossover (star location in Fig. 1b), an external out of plane magnetic field $\mu_0 H_{ext} \simeq 80$ μT stabilizes magnetic bubbles of $\simeq 1$ μm diameter (white dots under ITO in Fig. 1d, e, g, h). When a current is applied, magnetic bubbles drift in the same direction confirming their skyrmionic nature and their homo-chirality. We call them skyrmions in the following, since they share the same topology[39]. Owing to the low injected current density ($J \simeq 5 \times 10^9$ A/m$^2$, see Supplementary section II), some skyrmions remain motionless as they are probably pinned by defects.

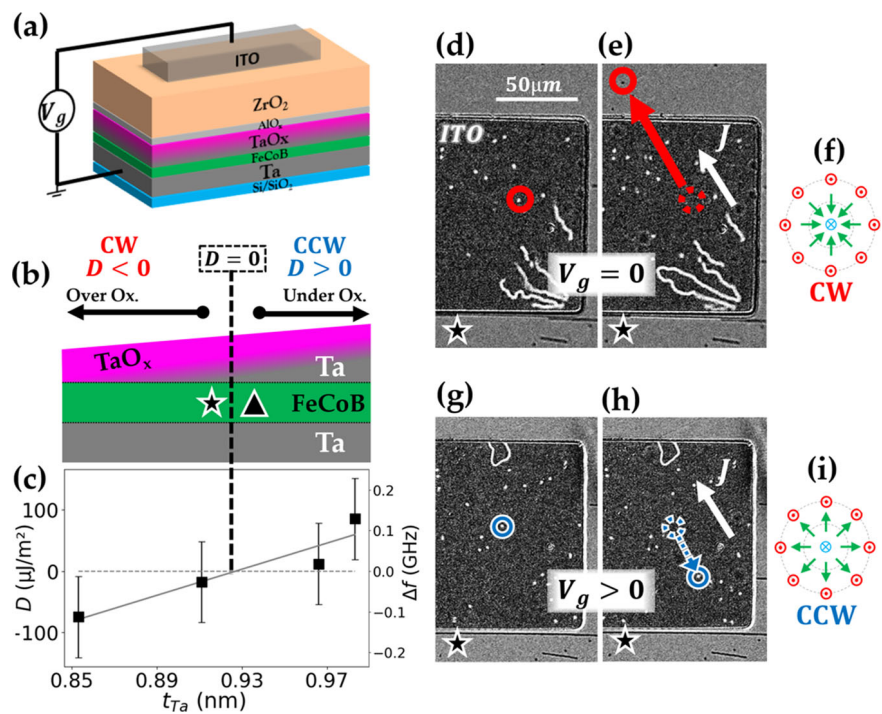

**Fig. 1 | Skyrmion chirality reversal. a** Schematic representation of the Ta/FeCoB/$TaO_x$ trilayer with additional $ZrO_2$ oxide and transparent ITO electrode for gate voltage application. **b** Schematic cross section of the sample: the top-Ta wedge induces an oxidation gradient at the top interface, leading to **c** a iDMI sign crossover as directly measured by BLS vs. top-Ta thickness. Error bars of ± 100 MHz on the frequency difference Δ*f*, represented on the right axis, are due to the setup. **d**, **e** CIM monitored during 4 s under p-MOKE microscope at the star location shown on **b** for zero gate voltage and **g**, **h** for $V_g$ = + 3.5 V, applied on ITO (the dark rectangular region). The in-plane current density ($J \simeq 5 \times 10^9$ A/$m^2$) is represented by the white arrow and the out of plane magnetic field is $\mu_0 H_{ext} \simeq 80$ μT. **d**, **e** In the initial state, skyrmions move in the direction of the current (encircled skyrmion moving along the red arrow), indicating CW chirality (*D* < 0), schematically represented in **f**. **g**, **h** Under the positive gate voltage, an inversion of the skyrmion motion occurs (encircled skyrmion moving along the blue arrow), indicating a CCW chirality (*D* > 0), as represented in **i**.

We then used transparent electrodes to directly observe the CIM and its inversion during or after the application of a gate voltage (See *Methods*). In the initial state ($V_g = 0$), the mobile skyrmions move along the current direction (speed $v_{0V} = 13.5 \pm 2$ μm/s at $J \simeq 5 \times 10^9$ A/m$^2$), which is expected for a Néel DW with a CW chirality (see red circles in Fig. 1d, e and Supplementary Video SV1). It is noteworthy that the effect of the thickness gradient on the skyrmion motion is negligible, indicating that current is the driving force (see Supplementary, section III). Besides, skyrmion Hall effect is expected to be negligible due to the small velocities in this regime of low current densities[40–42]. Furthermore, the continuous motion of skyrmions when crossing the edges of the electrode shows that the magnetic configuration is the same below and around the electrode.

Skyrmion CIM is then measured while applying a positive gate voltage on the electrode (Fig. 1g, h). We observe a progressive change: the skyrmion speed first decreases, then the motion direction inverts, typically after 90 s, and speed further increases and saturates. The CIM is now along the electron flow with $v_{+3.5V} = 3.2 \pm 2$ μm/s at $J \simeq 5 \times 10^9$ A/m$^2$ (see Supplementary, section IV, and Video SV2). Such inversion of motion is a signature of a transition from CW to CCW chirality, induced by an inversion of iDMI sign with the gate voltage. As expected, this CIM reversal is observed only below the ITO electrode, where the FeCoB/TaO$_x$ interface properties are modified by the gate voltage. This effect is reversible: switching the gate voltage to zero allows progressively recovering the as-grown CW skyrmion chirality, on the timescale of several minutes. Moreover, the chirality inversion is reproducible: skyrmions in Fig. 1 have previously undergone several chirality reversals. A more detailed analysis of skyrmions trajectories of Fig. 1 (d, e, g, h) and their inversion can be found in Supplementary, section IV, as well as the results of an experiment over a larger number of skyrmions, allowing extensive statistics.

Our experimental observations show that the gate voltage produces the same effect as a displacement along the Ta wedge from the region with $D < 0$ (star in Fig. 1b) to the region with $D > 0$ (triangle in Fig. 1b): starting from the region where skyrmions have CW chirality (as represented in Fig. 1f), a positive gate-voltage leads to a reversal to CCW chirality (as represented in Fig. 1i). Thus, a positive gate-voltage induces interfacial magnetic properties similar to those of a less oxidized interface. We may interpret this result either as a charge effect or as a migration of oxygen ions away from the interface. The former should produce an immediate effect whereas the latter is expected to be slower, progressive and possibly persistent. Since our measurements show that the reversal of the skyrmion motion occurs with a certain latency, we propose that the driving mechanism is ion migration. The positive gate voltage acts as a local and progressive reduction of the FeCoB/TaO$_x$ interface, that progressively decreases iDMI, eventually inverts its sign, thus triggering chirality reversal. Moreover, the recovery of the as-grown chirality when switching-off the gate voltage ($V_g = 0$) is consistent with the spontaneous progressive re-oxidation of the FM/MO$_x$ interface observed in similar materials with an equivalent timescale[43].

## Persistent and reversible control of chirality with gate voltage

Hereafter, we explore the chirality reversal process on labyrinthine domains (see Fig. 2) obtained by decreasing the external magnetic field to 30 μT in a region of the sample similar to the one of Fig. 1 (d, e, g, h). This magnetic configuration is more robust than skyrmions to small changes of magnetic parameters and magnetic field[11]. Here, we focus on the persistency of the effect of gate voltage on DW chirality. Thus, the current injection experiments, to probe the chirality, were performed after turning off the gate voltage.

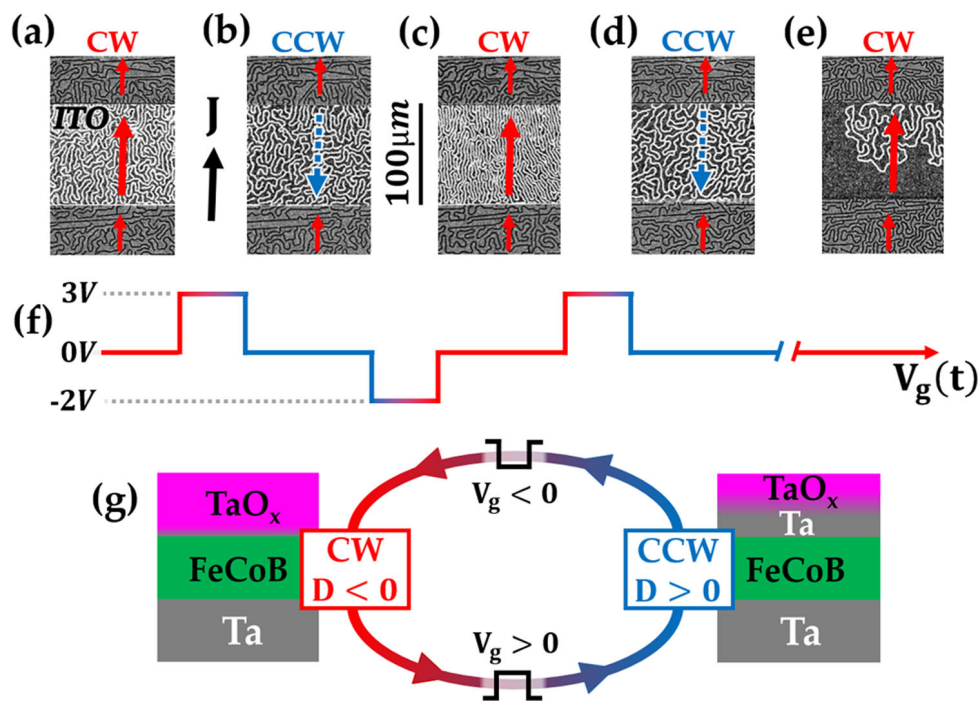

**Fig. 2 | Persistent and reversible chirality switch.** In the region close to iDMI sign inversion (star in Fig. 1b), the current density $J$ (black arrow) induces a motion of DWs (red/blue arrows for a motion along/opposite to the current density), as observed by p-MOKE microscopy after switching off the gate voltage.
**a–e** observation of DW motion under zero gate voltage and $\mu_0 H_{ext} \simeq 30$ μT, after sequential 90s-long voltage pulses. **f** Schematic representation of the applied voltage as a function of time. Initially **a** DWs have CW chirality; after a positive gate voltage pulse (**b**), chirality is reversed to CCW under the ITO electrode; after a negative gate voltage pulse (**c**), CW chirality is recovered ; after a positive gate voltage pulse (**d**), chirality has switched again to CCW; **e** after waiting ~2 h with zero gate voltage applied, the initial CW chirality is recovered. **g** Schematics of the effect of gate voltage pulses on interface oxidation, DW chirality and iDMI.

Figure 2 a shows the initial nearly demagnetized state with labyrinthine domains. The pattern of the labyrinthine domains is identical and continuous below and around the electrode. In the initial state, before gate voltage application, the DWs move in the same direction as the current, which is an indication of their CW chirality (see Fig. 2a and Supplementary Video SV3). After the application of a 90 s positive gate voltage pulse ($V_g$ = 3 V), the CIM of the DWs below the electrode is reversed (see Fig. 2b and Supplementary Video SV4), indicating a CCW chirality, which is due to an inversion of iDMI sign. This result is similar to the one obtained for skyrmions, except that the domain wall chirality is now probed ≃ 5 s after the gate voltage has been set to zero.

We further observed that a 90 s negative gate voltage pulse ($V_g$ = −2 V) restores the initial CW chirality (Fig. 2c and Supplementary Video SV5). A subsequent 90 s positive gate voltage pulse ($V_g$ = 3 V) once again switches towards CCW chirality (Fig. 2d and Supplementary Video SV6). Thus, chirality can be reversibly controlled by gate voltage and in a persistent way. It is reversed from CW to CCW (resp. from CCW to CW) with a positive (resp. negative) gate voltage, which we attribute to reduction (resp. oxidation) of the FeCoB/TaO$_x$ interface (see Fig. 2g).

When the gate voltage is set to zero, the reversed CCW DWs of Fig. 2d recover their initial CW chirality after about 2 h (Fig. 2e and Supplementary Video SV7), which is longer than the previous experiments on skyrmions for which only positive gate voltages were applied (see previous section). In this second experiment, negative voltages were applied, which induced further oxidization of the FM/MO$_x$ interface. This is thermodynamically favorable due to the affinity of metal for oxygen and induces a certain degree of irreversibility in some FM/MO$_x$ systems[21,31]. The slight difference in FeCoB thickness between skyrmion and stripe experiments may be at the origin of the slower recovery of magnetic properties after the application of a positive voltage. We suggest that the positive gate voltage drives oxygen ions from their equilibrium position into a metastable less oxidized state, in which they remain for some time after the gate voltage has been turned off. The existence of such a metastable state has been theoretically demonstrated at Fe/MgO interface, in the opposite case, ie. when interfacial oxygen is migrated towards the first Fe layer[44]. The slow recovery of the initial state, also reported in other studies[45,46], corresponds to a return to a metastable state where Ta naturally re-oxidizes. The timescale of this process is consistent with our hypothesis of oxygen migration, which is known to occur in TaO$_x$ and ZrO$_x$[32].

Finally, we have observed that chirality control can be achieved either starting from a negative iDMI (zone indicated by the star in Fig. 1b, see Fig. 2) or from a positive iDMI (zone indicated by the triangle in Fig. 1b, see Supplementary section V), by applying a gate voltage of appropriate polarity, as schematically represented in Fig. 2g.

**Stability of skyrmions under chirality reversal: analytical model and micromagnetic simulations**

The observed inversion of the skyrmion CIM under the application of a gate voltage is the signature of a transition between CW and CCW Néel skyrmions, which results from a iDMI sign inversion. In principle, this transition is possible without unraveling the spin texture since CW Néel, CCW Néel, and the expected intermediate Bloch skyrmion at zero iDMI share the same topology. However, even if this transformation is topologically allowed, it may affect the energetic stability of the skyrmion, in particular the stability of the Bloch skyrmion at zero iDMI. In the absence of stabilization by iDMI energy in thin films, only dipolar energy and out of plane external magnetic field may stabilize Bloch skyrmions[47].

To evaluate the stability of skyrmions during the application of a gate voltage, we have considered an analytical model describing the energy difference between an isolated skyrmion bubble and the uniform magnetic state[34] (see Methods). The magnetic parameters used in this analytical model and their variation under positive gate voltage are those extracted from experimental measurements (see Supplementary section VI).

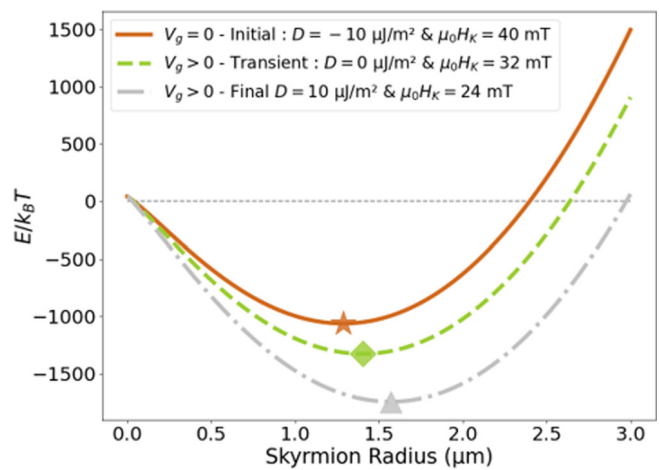

**Fig. 3 | Analytical model[34]: stability of skyrmions in FeCoB during iDMI inversion induced by the gate voltage.** Analytical calculation of energy difference (in units of $k_B T_{300K}$) between skyrmion and uniform state for FeCoB as a function of skyrmion radius. Solid orange, dashed green, and dash-dotted gray lines correspond, respectively, to negative, zero and positive iDMI, associated with a progressive anisotropy variation under the gate voltage, as experimentally measured. Owing to the small iDMI value in FeCoB, the slight change of equilibrium radius (depicted by symbols) is mostly due to the anisotropy variation.

The model predicts that for both non-zero and zero iDMI, a skyrmion is stable for diameters around 1.5 μm (see Fig. 3), close to the experimental values. Only a slight change of equilibrium diameter is expected, mostly due to the anisotropy variations under gate voltage, since our iDMI values, relatively small ($|D| \simeq 10$ μJ/m$^2$ interpolated from BLS measurements, see Supplementary section VI), contribute little to the total energy of the skyrmion. According to the model, it should be possible to reverse the chirality of a single skyrmion without annihilating it.

To better understand the mechanism of the chirality reversal, we carried out micromagnetic simulations[48] (see Methods). These enable investigating the chirality reversal mechanism at small dimensions inaccessible with the experimental setup of our study (Kerr microscope resolution ≃ 0.5 μm). The magnetic parameters used in this simulation (Co-based magnetic parameters[42], see Methods) lead to sub-micrometer size skyrmions, which are more relevant for applications. These small skyrmions cannot be described by the previous analytical calculations since their DW cannot be considered as infinitely thin with respect to skyrmion diameter. By contrast, they are more adapted to micromagnetic simulations as they require a reasonable number of cells. For small skyrmions, the iDMI contribution to the total energy is larger and we may thus wonder if their stability might be affected. The simulations were performed for iDMI value in the range [−0.5; 0.5] mJ/m$^2$. For each iDMI value, a magnetic skyrmion is stabilized, in particular for $D$ = 0 where a Bloch skyrmion is stable (see Fig. 4a–c). A typical electric field of $E$ = 1 V/nm, below the breakdown electric field in similar ZrO$_x$-based sample[43], is reasonable. A corresponding variation of iDMI of $\Delta D$ = 1 mJ/m$^2$ would require a iDMI variation efficiency under electric field $\beta_{iDMI} = \Delta D / E$ of 1000 fJ/(Vm). This is a proper order of magnitude for ionic effects[38] or in the case of ultrathin ferromagnets[27].

In the center of the DW, the angle $\xi$ between the in-plane magnetic moments and the radial direction, usually named helicity, evolves gradually from $|\xi|$ = 0 at $D$ = − 0.5 mJ/m$^2$ (CW Néel, see Fig. 4a) to $|\xi|$ = π at $D$ = 0.5 mJ/m$^2$ (CCW Néel, see Fig. 4c) via a $|\xi| = \frac{\pi}{2}$ Bloch skyrmion state at $D$ = 0 (see Fig. 4b). The radius variation between Néel and Bloch skyrmion (from 165 to 22 nm, see Fig. 4f) is much larger than in the analytical model prediction for FeCoB (see Fig. 3). This may be explained by a larger decrease of domain wall energy thanks to iDMI for the case of the Co-based sample (micromagnetic simulation).

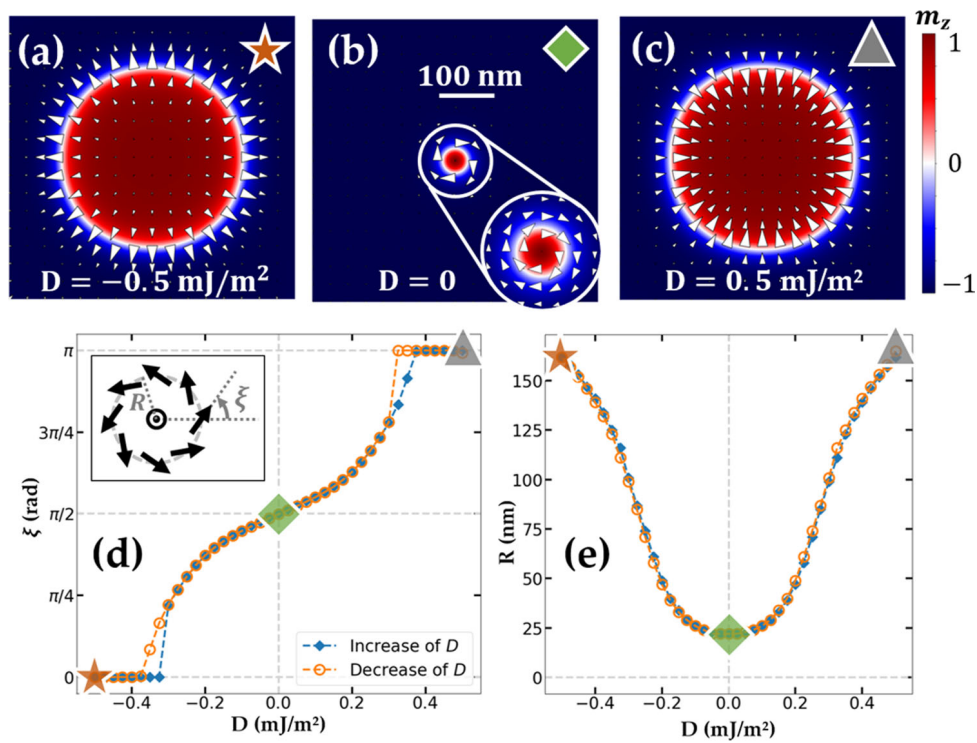

**Fig. 4 | Micromagnetic simulations of chirality switch.** Simulated stable states show a gradual transition between **a** CW Néel skyrmion at $D = -0.5$ mJ/m$^2$ and **c** CCW Néel skyrmion at $D = 0.5$ mJ/m$^2$ via **b** a stable Bloch skyrmion state at $D = 0$. **d** Angle $\xi$ between the in-plane magnetic moments and the radial direction of the magnetic moments at the domain-wall center and **e** radius of the skyrmion as a function of the iDMI value. The helicity and radius of the skyrmion corresponding to **a**–**c** images are shown, respectively, in **d**, **e** by the star, square and triangle symbols.

In these zero temperature simulations, the evolution of $\xi$ with iDMI (see Fig. 4d) presents a hysteretic behavior around $|D| = 0.35$ mJ/m$^2$. It corresponds to the beginning of the coherent rotation of the moments in the DW, selecting one of the two degenerated states leading to a CW or CCW Bloch skyrmion at $D = 0$ (see Supplementary, section VII).

## Discussion

Our observations of chirality reversal are due to gate voltage effect on iDMI. The effective iDMI originates from the two FeCoB interfaces in Ta/FeCoB/TaO$_x$. At the bottom Ta/FeCoB interface, the Fert-Levy mechanism[49,50] is at the origin of a small, negative iDMI contribution[19,51] (typically $-0.03$ mJ/m$^2$). The origin of iDMI for the top FeCoB/TaO$_x$ interface depends on oxidation state: (i) an underoxidized FeCoB/Ta top interface leads to a dominant Fert-Levy contribution to iDMI, with opposite sign with respect to the iDMI from the bottom interface and with larger magnitude due to intermixing[51]. This leads to a positive effective iDMI. (ii) By contrast, by gradually oxidizing this interface, the interfacial electric field at the origin of Rashba effect[52] is modified, resulting in a contribution to iDMI[18] found to be negative in this system[38] (in ref. 38, the convention for iDMI sign is opposite to the one used in the present article). Towards the more oxidized region, this Rashba contribution becomes dominant and determines the negative sign of the effective iDMI. Our experiments are done in the region where the contributions from the two interfaces almost cancel each other. Then, iDMI values are very small ($|D| \simeq 10$ μJ/m$^2$) and should result in hybrid Bloch-Néel domain walls, so called Dzyaloshinskii walls[9] (see Supplementary section VII, Fig. S9 (d, f, h, j)). However, the observed CIM parallel (resp. antiparallel) with the current density is the expected behavior of CW (resp. CCW) Néel DWs. In any case, an inversion of the iDMI sign leads to the inversion of the Néel component of the DW, which similarly leads to an inversion of the CIM direction.

It has been demonstrated that the perpendicular magnetic anisotropy and the iDMI have common origins, and are thus both sensitive to gate voltage[18] and oxidation[22]. Then, similarly to voltage control of magnetic anisotropy[28] (VCMA), an applied gate voltage can produce instantaneous reversible charge effects on iDMI[38] or persistent ones linked to ionic migration (see Supplementary of ref. 38). The relative contribution of charge and ionic effects on interfacial magnetism therefore depends on both the thickness of the ferromagnetic film and the oxidation of the FM/MO$_x$ interface. In the case of charge effects, the short screening length in metals (shorter than the FeCoB film thickness) would mainly modify charge distribution at the interface with the oxide. The addition of the applied electric field to the Rashba-field could reverse the total interfacial electric field, inducing an inversion of iDMI sign.

Nevertheless, we have shown that the effect of the gate voltage produces a similar effect as a displacement along the oxidation gradient. This is consistent with oxygen ion migration affecting the top FeCoB/TaO$_x$ interface. This migration induced by gate voltage may lead to a transition between negative Rashba and positive Fert-Levy contribution to iDMI. As we observed persistent effect on the timescale of minutes, we propose that oxygen ionic migration is the dominant mechanism observed in our study, as these ions are the mobile species in ZrO$_2$. The top-Ta nominal thickness varies by 0.015 nm over the region of interest (from triangle to star positions in Fig. 1c). Hence, the iDMI sign change is caused by the equivalent of 0.05 monolayer variation of the oxygen content at the FeCoB/TaO$_x$ interface. As previously discussed, a Ta thickness variation and a gate voltage application induce an equivalent slight change in oxidation. At a buried interface, such a small gate induced variation of a light element is challenging to observe using conventional microscopy techniques. However, monitoring the current-induced motion of chiral spin textures on a double wedge sample provides a powerful tool to access such small changes in composition.

Since iDMI may also be tuned by charge effects[38] that occur as sub-nanosecond timescale[53], we may envision an ultra fast switch of skyrmion chirality through a transient Bloch state. Furthermore, contrary to a current polarity inversion, which would invert similarly the motion direction of all skyrmions in the track, a gate voltage would provide a simple and local method to individually control skyrmions. Notably, their individual motion tuning can be fully exploited in race logic where information is stored in propagation time[54,55]. Eventually, due to the persistency of the effect, we may envision their use in artificial neural networks based on cross bar geometries[56] with multiple gates to dynamically and reversibly control the exact path of each input skyrmion. Besides, skyrmion motion along a track could be stopped by a Néel to Bloch transition using a gate voltage. This would enable an alternative realization of a skyrmion transistor hitherto proposed using VCMA[57,58]. Finally, this chirality switch offers a new degree of freedom, which could be used in reversible and programmable logic gates.

In summary, we have demonstrated a gate-voltage induced reversal of skyrmion chirality in Ta/FeCoB/TaO$_x$ through the inversion of their current-induced motion direction. Besides, we also observed a local, persistent and reversible chirality reversal of labyrinthine chiral domain walls by gate voltage. These reversals are due to an inversion of the iDMI sign and explained by the gate-controlled modification of the oxidation state at the ferromagnet/oxide interface. Micromagnetic simulations support the feasibility of a chirality reversal for submicronic skyrmions without annihilation. Such local and dynamical degree of freedom at the nanometer scale, controlled with voltages compatible with applications ($|V_g| \simeq 2–3$ V), would lay the foundations for efficient and multifunctional architectures involving magnetic skyrmions as information carriers.

## Methods
### Sample preparation
The base sample consists in a Ta(3)/FeCoB(1.1–1.3)/Ta(0.85–1) (nominal thicknesses in nm) crossed double wedge trilayer grown by magnetron sputtering on a thermally oxidized Si/SiO$_2$ wafer[59]. The top-Ta wedge was further oxidized in a treatment chamber (oxygen pressure 150 mbar for 10 s) thus leading to an oxidation gradient at the top interface (see Fig. 1b). In order to protect from further oxidation, a 0.5nm layer of Al was deposited and subsequently oxidized at air when taking the sample out of the sputtering machine. The final stack thus consists in a Ta(3)/FeCoB(1.1–1.3)/TaO$_x$(0.85–1)/AlO$_x$(0.5) (thicknesses in nm). Then, the sample was annealed (225 °C for 30 min) and a 20 nm ZrO$_2$ oxide was deposited by atomic layer deposition. This oxide layer acts as a dielectric and a ionic conductor. For this study, we restricted ourselves to constant ferromagnetic thicknesses ($t_{FeCoB} \simeq 1.2$ nm for skyrmion observation of Fig. 1 and $t_{FeCoB} \simeq 1.1$ nm labyrinthine pattern of Fig. 2), thus simplifying the sample to a single top-Ta wedge, as shown in Fig. 1b. The wedge of FeCoB was only used in the determination of some parameters (see Supplementary section VI). Finally, 70 nm transparent ITO electrodes were patterned by laser lithography. The size of the electrodes is 100 × 800 μm$^2$.

### Skyrmion observation, current-induced motion
The use of p-MOKE under transparent ITO electrodes allows probing the magnetization configuration both under and around the electrodes. Differential imaging is used in order to improve the contrast (the reference is the saturated magnetic state). Under ITO (resp. around it), black (resp. gray) regions correspond to magnetization pointing up, and white (resp. black) regions to magnetization pointing down. This variation of contrast might be explained by anti-reflecting effect from the ITO electrodes.

The skyrmion or labyrinthine phase (resp. in Figs. 1 and 2) is obtained by applying a constant perpendicular field ($\mu_0 H_{ext}$ of 80 and 30 μT, respectively, including Earth field), after saturating the magnetic state with a field of same polarity. Meanwhile, a current is injected via microbonded wires in the trilayer plane before applying any gate voltage in order to probe the initial chirality through the CIM direction. Then, the current is turned off and at this point, the measurement is different between skyrmions and labyrinthine domains.

For skyrmions, a gate voltage is continuously applied. During this time, chirality is regularly probed (every $\simeq 20$ s) by injecting current during sufficient time for the CIM to be measured.

For labyrinthine domains, voltage pulses are applied on the gate and CIM is measured after each pulse, i.e., when the voltage is turned off.

In differential imaging, mechanical drift can degrade the contrasts. To avoid it, the reference is renewed before each CIM measurement (short pulse of large magnetic field at which a new reference is taken). Finally, to illustrate the motion, a color-coded set of arrows indicates the CIM direction in Figs. 1 and 2.

### Direct iDMI measurements via Brillouin light scattering
The BLS setup used in this study consists of a linearly polarized LASER beam ($\lambda = 532$ nm) sent on the magnetic sample in the Damon-Esbach geometry (magnetization perpendicular to the light's incidence plane). The interaction of light with the spin waves can lead to the absorption or creation of a magnon, respectively called the Stokes and anti-Stokes event, that increases or decreases the frequency of the backscattered photons. The frequency spectrum of the backscattered photons in obtained by a tandem Fabry-Pérot interferometer. The frequency shift between the Stokes and anti-Stokes events is directly related to the iDMI constant D through $\Delta f = \frac{2\gamma}{\pi M_s} D k_{SW}$. In our configuration, the incident angle of light is 60°, inducing $k_{SW} = 20.45$/μm. Moreover, the magnetization $M_s = 1.54 \pm 0.06$ MA/m was extracted from VSM measurements and the gyromagnetic ratio has been taken to $\frac{\gamma}{2\pi} = 28.5$ GHz/T.

### Analytical model and micromagnetic simulations
The analytical model from ref. 34 estimates the energy difference between an individual skyrmion of radius $R$ and the saturated magnetic state. One must notice that this model is valid for $Q = \frac{K_u}{K_d} > 1$, where $K_u$ is the uniaxial anistropy and $K_d = \frac{1}{2}\mu_0 M_s^2$ is the shape anisotropy constant. In our case, we can extract from experimental parameters $Q = 1.02$, lying in the area of validity of the model. In this model, the energy difference between an individual skyrmion state and the saturated magnetization state is written as

$$\Delta E_{sb} = 2\pi R t \sigma_{DW} + 2\pi R^2 t \mu_0 M_S H_{ext} - \pi t^3 \mu_0 M_S^2 I(d) \quad (1)$$

where $\sigma_{DW}$ is the domain wall energy (containing exchange, anisotropy and iDMI energy), $t$ is the ferromagnetic layer thickness, $R$ is the bubble radius, $M_S$ is the saturation magnetization, $\mu_0 H_{ext}$ is the applied magnetic field and $I(d)$ is defined as

$$I(d) = -\frac{2}{3\pi} d \left[ d^2 + (1 - d^2) \frac{E(u^2)}{u} - \frac{K(u^2)}{u} \right] \quad (2)$$

where $d = \frac{2R}{t}$, $u = \frac{d^2}{1+d^2}$ and $E(u)$, $K(u)$ are elliptic integral defined as

$$E(u) = \int_0^{\pi/2} \sqrt{1 - u \sin^2(\alpha)} d\alpha \quad (3)$$

$$K(u) = \int_0^{\pi/2} \frac{d\alpha}{\sqrt{1 - u \sin^2(\alpha)}} \quad (4)$$

The parameters used in the analytical model are the FeCoB experimental parameters. The saturation magnetization $M_s = 1.54 \pm 0.06$ MA/m was measured with Vibrating Sample Magnetometer (VSM). The uniaxial anisotropy field $\mu_0 H_K = 40$ mT, and its variation under the application of a positive gate voltage was measured through hard-axis hysteresis loop (see Supplementary section VI). The FeCoB thickness $t_{FM}$ = 0.57 nm was used to take into account a magnetically dead layer, estimated with VSM measurements versus FeCoB nominal thickness. The exchange stiffness was fixed to $A_{ex}$ = 12 pJ/m[38]. Finally, an external magnetic field $\mu_0 H_{ext}$ = −750 µT was set in a direction opposite to the magnetization in the core of the skyrmion (destabilizing field).

Using micromagnetic simulations (Mumax3[48]), we computed an isolated skyrmion in an infinite magnetic thin film by computing the demagnetizing field from an infinite sample acting on the simulation region (See Supplementary section VII).

The magnetic parameters for the simulation are $M_s$ = 1.42 MA/m (magnetization), $t_{FM}$ = 0.9 nm (ferromagnetic thickness), $K_u$ = 1.27 × $10^6$ J/m$^3$ (uniaxial anisotropy), $\alpha$ = 0.37 (Gilbert damping), and $A_{ex}$ = 16 pJ/m (exchange stiffness). In addition to the dipolar field, an additional homogeneous magnetic field is set to $\mu_0 H_z$ = − 6 mT (destabilizing field). The simulation region is a 512 nm square, with a mesh size 1 nm × 1 nm × 0.9 nm.

First, we checked for the stabilization of a skyrmion with a positive iDMI value $D$ = 0.5 mJ/m$^2$ (Fig. 4c). Then, we decreased the iDMI value from $D$ = 0.5 mJ/m$^2$ to $D$ = − 0.5 mJ/m$^2$ by step of 5% and checked for the stabilization of skyrmion at each step. Finally, with the same procedure, we increased the iDMI value back to the initial $D$ = 0.5 mJ/m$^2$. In this simulation, the magnetic moments in the center of the DW experience a CCW in-plane rotation for both the decrease and the increase of iDMI. As a result, a CW Bloch skyrmion is observed for the decrease (see Fig. 4b) and a CCW for the increase of iDMI (see Supplementary section VII).

## Data availability
The MOKE data generated and analyzed in this study are provided in the Supplementary Information/Source Data file. The BLS data used in this study are available from the corresponding author on reasonable request.

## Code availability
The custom codes used during the current study are available from the corresponding author on reasonable request.

## Acknowledgements
The authors thank O. Fruchart, N. Rougemaille, M. Chshiev, L. Cagnon, and A. Masseboeuf for fruitful discussions. The authors acknowledge funding by the French ANR (contract ELECSPIN ANR-16-CE24-0018, contract ADMIS ANR-19-CE24-0019), DARPA TEE program through Grant No. MIPR HR0011831554. R.K. acknowledges funding by Nanosciences Foundation and by the People Programme (Marie Curie Actions) of the European Union's Seventh Framework Programme (FP7/2007-2013) under REA grant agreement no. PCOFUND-GA-2013-609102, through the PRESTIGE programme coordinated by Campus France.

## Author contributions
C.B. and H.B. designed the project. C.E.F., J.F., R.K., A.F., and S.A. fabricated the samples. D.O. collected and processed the BLS data. D.O., M.B., Y.R., and M.C. interpreted the BLS data. C.E.F. collected and processed the MOKE measurements with the help of I.J. C.E.F. performed the micromagnetic simulations with the help of L.B.P. C.E.F., H.B., C.B., L.R., and L.B.P. interpreted the micromagnetic simulations. H.B. performed the analytical model. C.E.F., J.F., S.P., L.R., O.B., G.G., C.B., and H.B. interpreted the MOKE results. C.E.F., C.B., and H.B. wrote the first draft of the manuscript. All authors discussed and commented on the first draft and final version of the manuscript.

## Competing interests
The authors declare no competing interests.

## Additional information
**Supplementary information** The online version contains supplementary material available at
https://doi.org/10.1038/s41467-022-32959-w.

**Correspondence** and requests for materials should be addressed to Hélène Béa.

**Peer review information** *Nature Communications* thanks the anonymous reviewer(s) for their contribution to the peer review of this work.

**Reprints and permission information** is available at
http://www.nature.com/reprints

**Publisher's note** Springer Nature remains neutral with regard to jurisdictional claims in published maps and institutional affiliations.